\def\@formatdoi#1{\url{}}
\documentclass[sigconf,authorversion]{acmart}

\acmDOI{}
\def\@formatdoi#1{\url{}}

\usepackage{ifthen}
\usepackage[normalem]{ulem} 

\newboolean{showcomments}
\setboolean{showcomments}{false} 

\newboolean{showreview}
\setboolean{showreview}{false} 

\ifthenelse{\boolean{showcomments}}
{
	\usepackage{mparhack}
	
	\newcommand{\nbb}[3]{
		\marginpar[\hspace*{-0.25cm}\parbox{35pt}{\tiny#1}]{\parbox{35pt}{\tiny#1}}
		{#2}
	}
	\newcommand{\modified}[1]{{\color{orange!80!black}#1}}
	\newcommand{\removed}[1]{{\color{red!90!black}\sout{#1}}}
	\newcommand{\added}[1]{{\color{green!49!black}#1}}
	\newcommand{\moved}[1]{{\color{blue!49!black}#1}}
	
	\newcommand{\rmodified}[2]{\nbb{#1}{\color{orange!80!black} #2}{orange!80!black}}
	\newcommand{\rremoved}[2]{\nbb{#1}{\color{red!90!black} \sout{#2}}{red!90!black}}
	\newcommand{\radded}[2]{\nbb{#1}{\color{green!49!black} #2}{green!49!black}}
	\newcommand{\rmoved}[2]{\nbb{#1}{\color{blue!49!black} #2}{teal}}
	
}
{
	\newcommand{\nbb}[3]{}
	\newcommand{\modified}[1]{#1}
	\newcommand{\added}[1]{#1}
	\newcommand{\removed}[1]{}
	\newcommand{\moved}[1]{#1}
	
	\newcommand{\rmodified}[2]{#2}
	\newcommand{\rremoved}[2]{}
	\newcommand{\radded}[2]{#2}
	\newcommand{\rmoved}[2]{#2}
}

\newcommand{\ext}[1]{}

\usepackage{reviews-style}

\usepackage{tabularx}
\usepackage{xltabular}
\usepackage{multicol}
\usepackage{multirow}

\usepackage[tikz]{ocgx2}
\usetikzlibrary{positioning,fit,arrows,shapes.geometric}
\usepackage{xcolor}

\AtBeginDocument{%
  \providecommand\BibTeX{{%
    \normalfont B\kern-0.5em{\scshape i\kern-0.25em b}\kern-0.8em\TeX}}}

\setcopyright{acmcopyright}
\copyrightyear{2022}
\acmYear{2022}

\acmConference[ESEM '22]{ESEM '22: ACM/IEEE International Symposium on Empirical Software Engineering and Measurement}{September 19--23, 2022}{Helsinki, Finland}
\acmBooktitle{ESEM '22:  ACM/IEEE International Symposium on Empirical Software Engineering and Measurement, September 19--23, 2022 Helsinki, Finland}
\acmPrice{15.00}
\acmISBN{978-1-4503-XXXX-X/22/09}

\begin{document}

\title{The Impact of Model Transformation Language Features on Quality Properties of MTLs: A Study Protocol}

\ifthenelse{\boolean{showreview}}{%
	        \pagenumbering{roman}%
	        \onecolumn

\begin{review-segment}{Description of Changes}
\begin{adjustwidth}{35pt}{35pt}

Dear Reviewers,
\bigskip\par

we thank you for the review of our registered report and are happy it was well received.

In the following, we discuss how we addressed your concerns with our report.
All changes throughout the manuscript are highlighted with different colors, according to whether the content was \textcolor{green!49!black}{added}, \textcolor{orange!80!black}{modified}, \textcolor{red!90!black}{\sout{deleted}}, or \textcolor{blue!49!black}{moved}.
For each major change we also reference the reviewer's comment that prompted the change (clicking the reference will move focus to the comment and our response).
Below you find the original review comments, interleaved with our responses.

\rule{\linewidth}{.75pt}
\vspace{-2cm}
\begin{reviewer-segment}{1}

The proposal is building on the authors’ previous work, in which they want to explore the impact of model transformation language features on quality properties of MTLs. The authors motivate the work properly with providing enough background about the concepts and the main technique to be used for their analysis. The technique is universal structural modelling which will be applied on the results obtained from answers of participants in a survey. Four RQs are explored, and the proposal falls under the two categories of confirmatory and exploratory studies. 
Some points are missing that I encourage authors to address in the extended version as follows.

\review{review-capabilities}{What are the MTL capabilities and MTL properties? What do they mean and what is their purpose? They each contain multiple variables that their definition is missing. The above-mentioned explanation helps the readers to understand the purpose and significance of the study.}
{Stefan}
{DONE}
{
We added Appendix~\ref{sec:descriptions} to give a structured overview over all capabilities and properties.
We put it as an appendix because they would disturb the reading flow.
}

\review{review-recruiting}{Some details about recruiting subjects are missing. Are they paid? }
{Stefan}
{DONE}
{
Participation in our study is completely voluntary and we do not plan to compensate subjects.
From our experience this is not necessary.
We have added this explanation to Section~\ref{sec:subjects}.
}

\review{review-survey-details}{Details about the survey are missing. Is this structured or semi-structured survey? How many questions exist? What is the avg time required for the participants to answer the questions? Is there any appraoch the authors are using to validate the survey results? Is there any alternate plan in case 80 participants are not found? Please add more details.}
{Stefan}
{DONE}
{
The survey is intended as a structured survey, has 30 questions and will take approximately 35 to 40 minutes to complete.
We added Section~\ref{sec:survey_plan} which contains all this information.

We intend to execute our analysis regardless of the final sample size but will critically assess the results in light of the actual sample size.
This explanation has been added to Section~\ref{sec:analysis_plan}.

The analysis results will be validated through tests suggested by the USM approach as explained in Section~\ref{sec:analysis_plan}.
We do not believe that there is a reasonable way to validate the responses of subjects directly.
}

\review{review-rq4-analysis}{The analysis plan for RQ4 is confusing. Please elaborate on this.}
{Stefan}
{DONE}
{
We agree that we do not properly explain the difference between explicitly hypothesised influences and influences that might be discovered during analysis.
We added clarification on this in Section~\ref{sec:analysis_plan} and rephrased the description of our process for \textbf{RQ4}.
}

\review{review-threats}{Is there any threat to the validity of the results?}
{Stefan}
{DONE}
{
There are validity threats to our results that stem from our subject sampling and the study limitations.
We added a discussion of the limitations and validity threats in Section~\ref{sec:threats}.
}

\review{review-difference-interviews}{I argue that although the motivation of the authors is based on their previous study and interview, and mentioning that this interview might not capture all effects, the base of the current proposal is on the same interview, when the authors assign the values of the likelihood of an interdependence between two variables. Please clarify this.}
{Stefan}
{DONE}
{
Assigning the likelihood values only provides a starting point for NEUSREL from which to start the internal calculations.
Because this starting point is necessary and the interviews do provide some confidence in the interdependencies we argue that using them as the basis for assigning is warranted.
Because there is still uncertainty about the completeness of interview responses we do not want exclude any interdependencies from the start.
We have extended the analysis plan in Section~\ref{sec:analysis_plan} with this explanation.
}

\review{review-significance-tests}{Other than USM, is there any other significance test applied in the study? According to authors, the USM quantifies the equations and values. How is the same technique used for the significance test? If so, is there any effect size calculated as well?}
{Stefan}
{DONE}
{
USM is a set of methods to confirm hypothesised influences and explore missed influences.
USM encompasses functionality to verify hypothesised influences, estimate influence and moderation weights between factors and a bootstrapping routine to calculate the statistical significance and effect sizes of the estimated results.
We use all methods defined in the USM approach to produce and validate our results.
We expanded our explanation of the approach in Section~\ref{sec:background} to make this more clear.
We also revised our reported analysis plan in Section~\ref{sec:analysis_plan} for this purpose.
}

\end{reviewer-segment}
\vspace{-1.75cm}
\begin{reviewer-segment}{2}
	
The article presents the design of a survey aiming to prove the existence of relations between model transformation languages features and their quality properties. The article aims to address a relevant problem, however, it appears to be still too immature. While I understand that a registered report should only report about the idea and its design, I also believe that such a design should be concrete and appliable for future experiments as is. In my opinion, this is not the case in this article, where more concrete details would be needed to improve the design. More detailed comments are as follows:

\review{review-hypotheses}{The hypotheses are not explicitly formalized. Section 4 reports a discussion about the hypotheses but they are never explicitly reported.}
{Stefan}
{DONE}
{
We agree that it can be confusing that all our hypotheses are only depicted implicitly in our structure model.
However due to the high number of hypotheses we believe this design to be necessary.
We added explicit explanation of this fact to Section~\ref{sec:rqs} and explained how all hypotheses can be inferred from the structure model.
Moreover, Section~\ref{sec:background} contains an example how a specific hypothesis is represented in a structure model.
}

\review{review-survey-details2}{More details about the survey structure should be reported. Section 7 only discusses the high-level sections of the survey but I would prefer to read a tentative structure including questions and other information such as the structure of single questions and the length of the survey.}
{Stefan}
{DONE}
{
We added Section~\ref{sec:survey_plan} which contains details on the structure and questions we intend to use in our survey.
The response options have been described in Section~\ref{sec:vars}.
}

\review{review-survey-implications}{What are the possible implications of this study? How can researchers and practitioners draw benefits from it?}
{Stefan}
{DONE}
{
Thank you for the suggestion.
We hope that the results can help researchers and practitioners make well-founded decisions on which languages to use, what factors to improve and what empirical studies, to validate the appropriateness of new techniques, should focus on.
We extended Section~\ref{sec:intro} with our vision for the use of our results.
}

\review{review-threats2}{The limitations of the designed study are only mentioned in the abstract. I suggest including a section where to report the major limitations and possible threats to the study validity.}
{Stefan}
{DONE}
{
We added a discussion of the limitations and validity threats in Section~\ref{sec:threats}.
}

\review{review-significance-tests2}{Which kind of test do you plan to run to verify your hypotheses? The execution plan (Section 7) includes both data extraction and data analysis but no additional details are provided.}
{Stefan}
{DONE}
{
We use the bootstrapping routine prescribed by USM to check the statistical significance of all hypothesised influences.
This is described in Section~\ref{sec:analysis_plan}.
We revised the section to make this more clear.
}

\review{review-structure}{Finally, I found some difficulties in reading the article. In my opinion, it needs some refactoring to ease the reading. For instance, subjects are reported before introducing the execution plan (and so the survey).}
{Stefan}
{DONE}
{
We agree that the structure can be confusing.
We moved our research questions (new Section~\ref{sec:rqs}) and execution plan (now Section~\ref{sec:execution_plan}) to the beginning of the paper and put the discussion of related work (now Section~\ref{sec:rw}) to the end.
We also added a short overview over the paper structure at the end of Section~\ref{sec:intro}.
We hope that this makes the structure easier to follow.
}

\end{reviewer-segment}

\bigskip

\rule{\linewidth}{.75pt}

In summary, we have considered all comments of the reviewers and updated our submission accordingly.

Yours sincerely,

Stefan Höppner and Matthias Tichy

\end{adjustwidth}
\end{review-segment}

%

\twocolumn

	        \pagenumbering{arabic}}{}%

\author{Stefan Höppner}
\orcid{0000-0001-7028-131X}
\affiliation{%
  \institution{Ulm University}
  \institution{Institute of Software Engineering and Programming Languages}
  \streetaddress{James-Franck-Ring 1}
  \city{Ulm}
  \country{Germany}
  \postcode{89081}
}
\email{stefan.hoeppner@uni-ulm.de}

\author{Matthias Tichy}
\affiliation{%
  \institution{Ulm University}
  \institution{Institute of Software Engineering and Programming Languages}
  \streetaddress{James-Franck-Ring 1}
  \city{Ulm}
  \country{Germany}
  \postcode{89081}
}
\email{matthias.tichy@uni-ulm.de}

\renewcommand{\shortauthors}{S. Höppner and M. Tichy}

\begin{abstract}
  \textbf{Background}:
  Dedicated model transformation languages are claimed to provide many benefits over the use of general purpose languages for developing model transformations.
  However, the actual advantages and disadvantages associated with the use of model transformation languages are poorly understood empirically.
  There is little knowledge over what advantages and disadvantages hold in which cases and where they originate from.
  In a prior interview study, we elicited expert opinions on what advantages result from what factors surrounding model transformation languages as well as a number of moderating factors that moderate the influence.
  
\noindent  \textbf{Objective}:
  We aim to quantitatively asses the interview results to confirm or reject the influences and moderation effects posed by different factors and to gain insights into how valuable different factors are to the discussion.
  
\noindent  \textbf{Method}: 
  We gather data on the factors and quality attributes using an online survey.
  To analyse the data and examine the hypothesised influences and moderations we use universal structure modelling based on a structural equation model.
  Universal structure modelling will produce significance values and path coefficients for each hypothesised and modelled interdependence between factors and quality attributes that can be used to confirm or reject correlation and to weigh the strength of influence present.
  
\noindent  \textbf{Limitations}:
  Due to the complexity and abstractness of the concepts under investigation, a measurement via reflective or formative indicators is not possible.
  Instead participants are queried about their assessment of concepts through a single item question.
  We further assume that positive and negative effects of a feature are more prominent if the feature is used more frequently.
\end{abstract}




\maketitle

\section{Introduction}
\label{sec:intro}

Model transformations lie at the centre of the model driven engineering (MDE) approach~\cite{Sendall2003}.
They are used to manipulate and evolve models~\cite{metzger2005systematic}, to derive artefacts like source code or documentation, to simulate system behaviour and to analyse system aspects~\cite{Schmidt2006}.

To aid development of model transformations numerous model transformation languages (MTLs) of different form, aim and syntax~\cite{kahani2019survey} have been proposed.
Using MTLs for the development of model transformations is claimed to have many benefits compared to using general purpose languages (GPLs)~\cite{Goetz2020}.
Claims include, but are not limited to, better \textit{comprehensibility}, \textit{productivity} and \textit{maintainability} as well as easier \textit{development} in general~\cite{Goetz2020}.
The emergence of such claims can, in part, be attributed to the advantages that are ascribed to domain specific languages (DSLs)~\cite{Hermans2009,Johannes2009}.

Whether these advantages exist and particularly where they arise from is, however, still uncertain, as discussed in our prior systematic literature review~\cite{Goetz2020}.
This makes it hard to convincingly argue MTLs over GPLs.
Especially considering recent GPL advancements, like Java Streams, C\# LINQ or advanced pattern matching syntax, that help reduce boilerplate code~\cite{Hoeppner2021} and have put them back into the discussion for transformation development.
A recent community discussion held at the 12th edition for the International Conference on Model Transformations (ICMT'19) even acknowledges GPLs as suitable contenders~\cite{Burgueno2019}.
The few empirical studies that exist on this topic also provide mixed and limited results.
Hebig et al. found no direct advantage for the development of transformations, but they did find an advantage for the comprehensibility of the code in their limited setup~\cite{Hebig2018}.
A study conducted by us, on the other hand, found that certain use cases favour the use of MTLs, while in others the versatility of GPLs may be more appropriate~\cite{Hoeppner2021}.
Overall there is a knowledge gap pertaining to what the exact benefits are, how strong their impact is, where they originate from and when they are valid.

To alleviate this gap we developed and executed an interview study with 56 experts from research and industry to discuss the topic of advantages and disadvantages~\cite{Hoeppner2022}.
During the interviews participants were queried about their views on the advantages and disadvantages of model transformation languages and the origins thereof.
Results from the interviews point towards three main-areas, namely \textit{general purpose languages capabilities}, \textit{model transformation languages capabilities} and \textit{tooling}, relevant to the discussion.
Interviewees provided information on which claimed MTL properties are influenced by which sub-areas and why.
Interview participants also provided us with insights on moderation effects on these interdependencies caused by different \textit{use-cases}, \textit{skill \& experience levels} of users and \textit{choice of transformation language}.

The results of our interview study are, however, purely qualitative and are therefore limited in their informative value.
They do not provide indication on the strength of influence between the involved variables.
Moreover, they are based on the views of a limited number of participants.
It is not clear if the extracted influence relationships are complete or whether they withstand community scrutiny.
As a result, they only represent an initial data set that requires a quantitative and detailed analysis.

In this registered report, we detail a study protocol to \textit{confirm or deny} the interdependencies hypothesised from our interview results.
We further intend to \textit{quantify the influence strengths} and to \textit{gain insights into moderation effects} suggested during the interviews.
Lastly, we aim to explore whether there are interdependencies between factors and quality properties that were missed by our interviewees.

This study will focus on the effects of \textit{MTL capabilities} (namely Bidirectionality, Incrementality, Mappings, Model Management, Model Navigation, Model Traversal, Pattern Matching, Reuse Mechanisms and Traceability) on \textit{MTL properties} (namely Comprehensibility, Ease of Writing, Expressiveness, Productivity, Maintainability and \modified{Reusability} and Tool Support) in the context of \textit{bidirectional} or \textit{incremental use-cases}, \textit{language choice}, \textit{language skills}, \textit{experience}, \textit{meta-model sanity}, \textit{model size}, \textit{meta-model size}, and \textit{semantic gap between input and output}.
Further studies can follow the same approach and focus on different areas.
Short descriptions for all MTL capabilities and MTL properties can be found in Table~\ref{tbl:cap_descriptions}~\&~\ref{tbl:prop_descriptions} in Appendix~\ref{sec:descriptions}.
More thorough explanations can be found in our previous works \cite{Goetz2020,Hoeppner2022}.

We plan to use an \textit{online survey} to query users of MTLs from research and industry about the amount they use different language features, their perception of qualitative properties of their transformations and demographic data surrounding use-case, skills \& experience and used languages.
The responses are used as input for universal structure modelling (USM)~\cite{buckler2008identifying} based on a structural equation model developed from the interview responses.
USM is a structural equation modelling approach designed on the premise that purely theory-testing approaches can not satisfactorily cope with the intricacies of many modelled issues~\cite{buckler2008identifying}.
It is used to estimate the influence weights between all variables within the structural equation model which can be used to gain insights about the influence and moderation strengths and thus help confirm or reject the hypothesised influences.
It can also help discover interdependencies between variables of the model that have not been actively modelled, to aid in developing a more complete model.

USM is chosen over its structural equation modelling alternatives due to it being able to better handle uncertainty about the completeness of the hypothesis system under investigation, it having more capabilities to analyse moderation effects and the ability to investigate non linear correlations~\cite{Weiber2021}.

\radded{\ref{review-survey-implications}}{
With this work we aim to provide researchers and practitioners with a fundamental model on influences and effects on the properties of model transformation languages.
We hope that the results can help researchers and practitioners make well-founded decisions on which languages to use, what MTL capabilities to improve and what empirical studies, to validate the appropriateness of new techniques, should focus on.
The influence weights of MTL capabilities can be used by researchers to focus their research on the most relevant capabilities for the properties they are interested in improving.
It can also help design appropriate empirical studies to evaluate the effectiveness of new techniques, e.g. for model management, by revealing which properties are most impacted by them.
Similarly, it can help practitioners in their decision making process when deciding on the use of MTLs.
It can allow them to identify relevant parts of their transformations to evaluate, depending on what properties they are most concerned about.
}

\radded{\ref{review-structure}}{
The structure of this registered report is as follows:
Section~\ref{sec:background} gives a short overview over model transformation languages and introduces universal structure modelling.
Section~\ref{sec:rqs} describes our research questions and hypotheses.
Afterwards in Section~\ref{sec:execution_plan} the execution plan for our study is introduced.
Section~\ref{sec:vars} details the variables involved in our study while Section~\ref{sec:survey_plan} details the structure of the survey and an excerpt of the questions used.
Section~\ref{sec:subjects} describes our intended target subjects and sampling methodology.
In Section~\ref{sec:analysis_plan} we describe how we plan to analyse the data produced in the online survey and Section~\ref{sec:threats} discusses threats to validity.
Lastly, in Section~\ref{sec:rw} we present and discuss related work.
}
\section{Background}
\label{sec:background}

\begin{figure*}[ht]
	\includegraphics[scale=0.65]{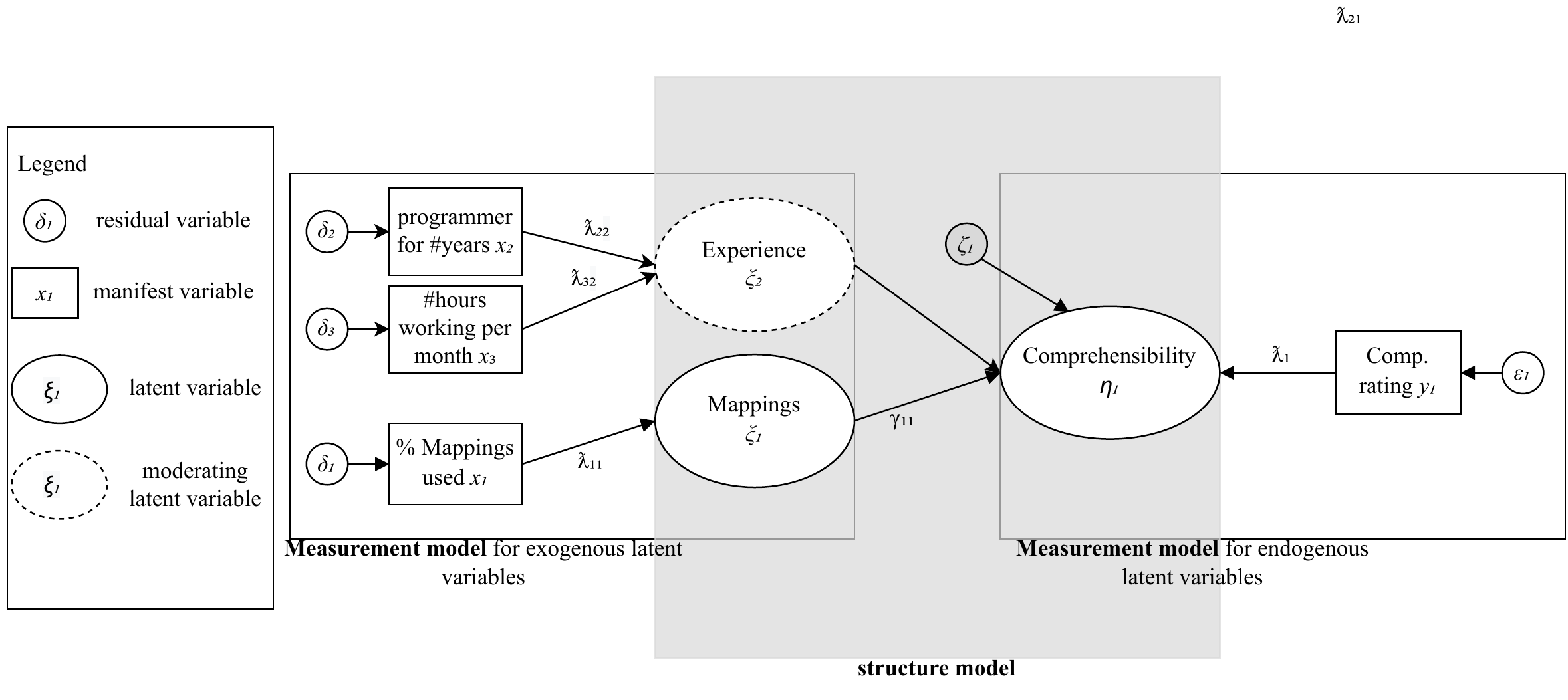}
	\caption{The makeup of a structural equation model.}~\label{fig:sample_SEM}
\end{figure*}


\textbf{Model transformation languages} have been developed to provide domain specific languages for developing model transformations.
Similar to general purpose languages, MTLs follow different paradigms and have various forms.
The axes along which MTLs can be distinguished have been explored in several studies, such as by Mens et al.~\cite{Mens2006}, Czarnecki et al.~\cite{Czarnecki2006} and more recently Kahani et al.~\cite{kahani2019survey}.

Although many distinctions exist, the fundamental goal and functionality provided by model transformation languages stays the same throughout all of them.
Model transformation languages are \textit{``language[s] that provide a set of constructs for explicitly expressing, composing, and applying transformations''}~\cite{Sendall2003}.

\noindent\modified{\textbf{Structural equation modelling (SEM)} is an approach used for confirmatory factor analysis~\cite{10.1145/3469888}.
It} defines a set of methods used to \textit{``investigate complex relationship structures between variables and allows for quantitative estimates of interdependencies thereof.
Its goal is it to map the a-priori formulated cause-effect relationships into a linear system of equations and to estimate the model parameters in such a way that the initial data, collected for the variables, are reproduced as well as possible''}~\cite{Weiber2021}.

Structural equation modelling distinguishes between two sets of variables \textit{manifest} and \textit{latent}.
\textit{Manifest} variables are variables that are empirically measured and \textit{latent} variables describe theoretical constructs that are hypothesised to interact with each other.
Latent variables are further divided into \textit{exogenous} or independent and endogenous or \textit{dependent} variables.

So called structural equation models, a sample of which can be seen in Figure~\ref{fig:sample_SEM}, comprised of manifest and latent variables, form the heart of analysis.
They are made up of three connected sub-models.
The \textit{structure model}, the \textit{measurement model} of the exogenous latent variables and the \textit{measurement model} of the endogenous latent variables.

The \textit{structure model} defines all hypothesised interactions between exogenous ($\xi_{exID}$) and endogenous ($\eta_{endID}$) latent variables.
Each exogenous variable is linked, by arrow, to all endogenous variables that are presumed to be influenced by it.
Each of these connections is given a variable ($\gamma_{exID\_endID}$) that measures the influence strength.
If an exogenous variable \textit{moderates} the influences on a endogenous variable, the exogenous variable is depicted with a dashed outline and connected to all endogenous variables that are moderated by it\footnote{To illustrate moderation, arrows are usually shown from the moderating exogenous variable to the arrow representing the moderated influence , i.e., an arrow between an exogenous variable and an endogenous variable.
However our illustration deviates from this due to the size and makeup of our hypothesis system.
For standard representations please refer to foundation literature such as Weiber and Mühlhaus~\cite{Weiber2021}.}.
For each moderated influence a separate variable of the form $\gamma_{exID\_endID\_modEndID}$ is assigned.
In addition, an residual (or error) variable is appended to each endogenous latent variable to represent the influence of variables not represented in the model.

Figure~\ref{fig:sample_SEM} shows an example structure equation model model for the hypothesis that \textit{``Mappings help with the comprehensibility of transformations, depending on the developers experience.''}.
The structure model seen at the centre of the figure, is comprised of the exogenous latent variable $\xi_1$ (\textit{Mappings}), the moderating exogenous variable $\xi_2$ (\textit{Experience}), the endogenous latent variable $\eta_1$ (\textit{Comprehensibility}), a presumed influence of \textit{Mappings} on \textit{Comprehensibility} via $\gamma_{11}$ and the error variable $\zeta_1$.
Lastly the model also contains a moderation of Experience on all influences of Comprehensibility.
As described earlier, this moderation effect is assigned the variable $\gamma_{11\_2}$.
The moderation variables are not depicted in our structure models because of their high number and visibility issues caused by it.

The \textit{measurement model} of the exogenous latent variables reflects the relationships between all exogenous latent variables and their associated manifest variables.
Each manifest variable is linked, by arrow, to all exogenous latent variables that are measured through it.
Each of these connections is given a variable that measures the indication strength of the manifest variable for the latent variable.
Additionally, an error variable for each manifest variable is introduced that represents measurement errors.
In Figure~\ref{fig:sample_SEM} the measurement model for exogenous latent variables, seen at the left of the figure, is comprised of the exogenous latent variables $\xi_1$ (\textit{Mappings}) and $\xi_2$ (\textit{Experience}), the manifest variables $x_1$ (\textit{\% of code using Mappings}), $x_2$ (\textit{number of years a person has been a programmer}) and $x_3$ (\textit{number of hours per month spent developing transformations}) their measurement accuracy for Mapping usage $\lambda_{11}$ and their measurement accuracy for Experience $\lambda_{22}$ and and $\lambda_{32}$ and the associated measurement error $\delta_1$ and $\delta_2$ and $\delta_3$.

The \textit{measurement model} of the endogenous latent variables reflects the relationships between all endogenous latent variables and their associated manifest variables.
It is structured the same way as the \textit{measurement model} of the exogenous latent variables.
In Figure~\ref{fig:sample_SEM} it is shown on the right of the figure.

Given a structural equation model and measurements for manifest variables the SEM approach calls for estimating the influence weights and latent variables within the models.
This is done in alternation for the measurement models and the structure model until a predefined quality criterion is reached.
Traditional methods (covariance-based structural equation modeling \& partial least squares) use different mathematical approaches such as maximum-likelihood estimation or least squares~\cite{Weiber2021} to estimate influence weights.

\noindent\rmoved{\ref{review-significance-tests}}{\textbf{Universal structure modelling (USM)} evolved from structural equation modelling as an alternative to the commonly used methods with the aim of coping with effects that are not initially theorized and non-linearity of influences~\cite{buckler2008identifying}.}
\radded{\ref{review-significance-tests}}{
To do so, USM encompasses functionality to verify hypothesised influences, estimate influence and moderation weights between factors and methods for calculating the statistical significance and effect sizes of the estimated results.
}
\rmoved{\ref{review-significance-tests}}{Similar to the partial least square approach, USM also uses least squares for estimating the measurement models but relies on a Bayesian neural network, that is composed of a multilayer perceptron architecture, to estimate the structure model~\cite{buckler2008identifying,Weiber2021}.
This has the advantage of reliably handling non-linear interdependencies and moderation caused, for example, through co-founding variables within the model.
To translate the neural network results into a understandable representations, USM introduces a post-processing stage where the regression functions from the network are analysed to extract properties of the paths, such as their influence strength, the form of the non-linearity or the form of interactions~\cite{Weiber2021}.}

Using USM instead of traditional structural equation modelling approaches is suggested for studies where there are still uncertainties about the completeness of the underlying hypotheses system and for exploring non-linearity in the influences~\cite{Weiber2021,buckler2008identifying}.
Moreover its use of a neural network also reduces the requirements for the scale levels of data thus allowing the introduction of categorical variables in addition to metric variables~\cite{Weiber2021}.
\moved{\section{Hypotheses and Research Questions}}
\label{sec:rqs}
\rmoved{\ref{review-structure}}{}

The goal of our study is to provide first results on the influence strengths of interdependences between model transformation language \textit{capabilities} and claimed \textit{quality properties} as perceived by users, as well as insights into the strength of moderation expressed by \textit{contextual properties}.
The study is structured around the hypothesised interdependencies between these variables, and their more detailed breakdown, extracted from the interview results.
Each presumed influence of a \textit{MTL capability} on a \textit{MTL property} forms one hypothesis which is to be examined and weighted in our study.
All hypotheses are extended with an assumption of moderation through the context variables as a result of the deliberations from the interview study.
The system of hypotheses that arises from these deliberations can be represented in a structure model, which forms the basis for our study.
The structure model is depicted in Figure~\ref{fig:structure_model}.
The model shows exogenous variables on the left and right and endogenous variables at the centre.
Exogenous variables depicted in a ellipse with a dashed outline constitute the hypothesised moderating variables.

\radded{\ref{review-hypotheses}}{
All hypotheses investigated in our study are of the form: \textit{``<MTL Property> is (positively or negatively) influenced by <MTL Capability>''}.
They are represented by arrows from exogenous variables on the left of Figure~\ref{fig:structure_model} to endogenous variable at the centre.
A moderation on the hypothesised influence is also assumed from all exogenous variables on the right of the figure connected to the considered endogenous variable.
In total we investigate 31 hypothesised influences, i.e. the number of outgoing arrows from the exogenous variables on the left of Figure~\ref{fig:structure_model}.
}

\tikzstyle{factor} = [ellipse, draw, node distance=2cm, text width=6em, text centered, minimum height=5em]
\tikzstyle{mod-factor} = [ellipse, draw, node distance=2cm, text width=6em, text centered, minimum height=5em,dashed]
\tikzstyle{property} = [ellipse, rounded corners, draw, node distance=3.25cm, text width=6em, text centered, align=center, minimum height=5em,fill=white]

\tikzstyle{error} = [circle, draw, text centered, xshift=4em, yshift=1em]

\tikzstyle{l} = [draw, -latex']

\pgfdeclarelayer{background}
\pgfsetlayers{background,main}

\renewcommand{\labelitemi}{$\bullet$}

\begin{figure*}
	\begin{tikzpicture}
		\node[factor, actions ocg={}{reset-b bx2comp bx2eow bx2exp bx2mtb bx2pro comp eow exp mtb pro}{inc2comp inc2eow inc2exp map2comp map2eow map2exp map2mtb map2reu man2comp man2eow man2exp nav2comp nav2eow nav2exp trv2comp trv2eow trv2exp trv2pro pm2comp pm2exp pm2pro rm2reu trc2comp trc2eow trc2exp trc2pro ts reu}] (factor-bx) at (0,-2) {Bidirectionality $\xi_1$};
		\node[factor,below of=factor-bx,actions ocg={}{reset-b inc2comp inc2eow inc2exp comp eow exp}{bx2comp bx2eow bx2exp bx2mtb bx2pro map2comp map2eow map2exp map2mtb map2reu man2comp man2eow man2exp nav2comp nav2eow nav2exp trv2comp trv2eow trv2exp trv2pro pm2comp pm2exp pm2pro rm2reu trc2comp trc2eow trc2exp trc2pro ts mtb pro reu}] (factor-inc) {Incrementality $\xi_2$};
		\node[factor,below of=factor-inc,actions ocg={}{reset-b map2comp map2eow map2exp map2mtb map2reu comp eow exp mtb reu}{bx2comp bx2eow bx2exp bx2mtb bx2pro inc2comp inc2eow inc2exp man2comp man2eow man2exp nav2comp nav2eow nav2exp trv2comp trv2eow trv2exp trv2pro pm2comp pm2exp pm2pro rm2reu trc2comp trc2eow trc2exp trc2pro ts pro}] (factor-map) {Mappings $\xi_3$};
		\node[factor,below of=factor-map,actions ocg={}{reset-b man2comp man2eow man2exp comp eow exp}{bx2comp bx2eow bx2exp bx2mtb bx2pro inc2comp inc2eow inc2exp map2comp map2eow map2exp map2mtb map2reu nav2comp nav2eow nav2exp trv2comp trv2eow trv2exp trv2pro pm2comp pm2exp pm2pro rm2reu trc2comp trc2eow trc2exp trc2pro ts mtb pro reu}] (factor-man) {Model Management $\xi_4$};
		\node[factor,below of=factor-man,actions ocg={}{reset-b nav2comp nav2eow nav2exp comp eow exp}{bx2comp bx2eow bx2exp bx2mtb bx2pro inc2comp inc2eow inc2exp map2comp map2eow map2exp map2mtb map2reu man2comp man2eow man2exp trv2comp trv2eow trv2exp trv2pro pm2comp pm2exp pm2pro rm2reu trc2comp trc2eow trc2exp trc2pro ts mtb pro reu}] (factor-nav) {Model Navigation $\xi_5$};
		\node[factor,below of=factor-nav,actions ocg={}{reset-b trv2comp trv2eow trv2exp trv2pro comp eow exp pro}{bx2comp bx2eow bx2exp bx2mtb bx2pro inc2comp inc2eow inc2exp map2comp map2eow map2exp map2mtb map2reu man2comp man2eow man2exp nav2comp nav2eow nav2exp pm2comp pm2exp pm2pro rm2reu trc2comp trc2eow trc2exp trc2pro ts mtb reu}] (factor-trv) {Model Traversal $\xi_6$};
		\node[factor,below of=factor-trv,actions ocg={}{reset-b pm2comp pm2exp pm2pro comp exp pro}{bx2comp bx2eow bx2exp bx2mtb bx2pro inc2comp inc2eow inc2exp map2comp map2eow map2exp map2mtb map2reu man2comp man2eow man2exp nav2comp nav2eow nav2exp trv2comp trv2eow trv2exp trv2pro rm2reu trc2comp trc2eow trc2exp trc2pro eow ts mtb reu}] (factor-pm) {Pattern Matching $\xi_7$};
		\node[factor,below of=factor-pm,actions ocg={}{reset-b rm2reu reu}{bx2comp bx2eow bx2exp bx2mtb bx2pro inc2comp inc2eow inc2exp map2comp map2eow map2exp map2mtb map2reu man2comp man2eow man2exp nav2comp nav2eow nav2exp trv2comp trv2eow trv2exp trv2pro pm2comp pm2exp pm2pro trc2comp trc2eow trc2exp trc2pro comp eow exp ts mtb pro}] (factor-rm) {Reuse Mechanisms $\xi_8$};
		\node[factor,below of=factor-rm,actions ocg={}{reset-b trc2comp trc2eow trc2exp trc2pro comp eow exp pro}{bx2comp bx2eow bx2exp bx2mtb bx2pro inc2comp inc2eow inc2exp map2comp map2eow map2exp map2mtb map2reu man2comp man2eow man2exp nav2comp nav2eow nav2exp trv2comp trv2eow trv2exp trv2pro pm2comp pm2exp pm2pro rm2reu ts mtb reu}] (factor-trc) {Traceability $\xi_9$};
		
		\node[mod-factor] (mfactor-mms) at (15,0) {Meta-model Size $\xi_{13}$};
		\node[mod-factor,below of=mfactor-mms] (mfactor-ms) {Model Size $\xi_{14}$};
		\node[mod-factor,below of=mfactor-ms] (mfactor-ts) {Transformation Size$\xi_{15}$};
		\node[mod-factor,below of=mfactor-ts] (mfactor-bus) {Bidirectional Use$\xi_{18}$};
		\node[mod-factor,below of=mfactor-bus] (mfactor-mtl) {Language Choice $\xi_{10}$};
		\node[mod-factor,below of=mfactor-mtl] (mfactor-lsk) {Language Skills $\xi_{11}$};
		\node[mod-factor,below of=mfactor-lsk] (mfactor-exp) {Experience $\xi_{12}$};
		\node[mod-factor,below of=mfactor-exp] (mfactor-io) {I/O Semantic gap $\xi_{16}$};
		\node[mod-factor,below of=mfactor-io] (mfactor-msa) {Meta-model sanity $\xi_{17}$};
		\node[mod-factor,below of=mfactor-msa] (mfactor-ius) {Incremental Use $\xi_{19}$};
		
		\node[property,actions ocg={}{reset-b dom2comp bx2comp inc2comp map2comp trc2comp trv2comp pm2comp nav2comp man2comp comp}{bx2eow bx2exp bx2mtb bx2pro inc2eow inc2exp map2eow map2exp map2mtb map2reu man2eow man2exp nav2eow nav2exp trv2eow trv2exp trv2pro pm2exp pm2pro rm2reu trc2eow trc2exp trc2pro eow exp ts mtb pro reu}] (prop-comp) at (7.5,-0.5) {Comprehensi-bility $\eta_1$};
		
		\node[error, above of=prop-comp] (comp-error) {$\zeta_1$};
		\path[l] (comp-error.south west) -- (prop-comp.45);
		
		\node[property,below of=prop-comp,actions ocg={}{reset-b bx2eow inc2eow man2eow nav2eow trv2eow trc2eow lea2eow use2eow ide2eow eow}{bx2comp bx2exp bx2mtb bx2pro inc2comp inc2exp map2comp map2exp map2mtb map2reu man2comp man2exp nav2comp nav2exp trv2comp trv2exp trv2pro pm2comp pm2exp pm2pro rm2reu trc2comp trc2exp trc2pro comp exp ts mtb pro reu}] (prop-eow) {Ease of Writing $\eta_2$};
		
		\node[error, above of=prop-eow] (eow-error) {$\zeta_2$};
		\path[l] (eow-error.south west) -- (prop-eow.45);
		
		\node[property,below of=prop-eow,actions ocg={}{reset-b bx2exp inc2exp map2exp man2exp nav2exp trv2exp pm2exp trc2exp exp}{bx2comp bx2eow bx2mtb bx2pro inc2comp inc2eow map2comp map2eow map2mtb map2reu man2comp man2eow nav2comp nav2eow trv2comp trv2eow trv2pro pm2comp pm2pro rm2reu trc2comp trc2eow trc2pro comp eow ts mtb pro reu}] (prop-ex) {Expressiveness $\eta_3$};
		
		\node[error, above of=prop-ex] (exp-error) {$\zeta_3$};
		\path[l] (exp-error.south west) -- (prop-ex.45);
		
		\node[property,below of=prop-ex,actions ocg={}{reset-b debug2ts mat2ts lea2ts val2ts use2ts ide2ts int2ts cre2ts ana2ts eco2ts rep2ts awa2ts ts}{bx2comp bx2eow bx2exp bx2mtb bx2pro inc2comp inc2eow inc2exp map2comp map2eow map2exp map2mtb map2reu man2comp man2eow man2exp nav2comp nav2eow nav2exp trv2comp trv2eow trv2exp trv2pro pm2comp pm2exp pm2pro rm2reu trc2comp trc2eow trc2exp trc2pro comp eow exp mtb pro reu}] (prop-ts) {Tool Support $\eta_4$};
		
		\node[error, above of=prop-ts] (ts-error) {$\zeta_4$};
		\path[l] (ts-error.south west) -- (prop-ts.45);
		
		\node[property,below of=prop-ts,actions ocg={}{reset-b bx2mtb map2mtb ide2mtb eco2mtb mtb}{bx2comp bx2eow bx2exp bx2pro inc2comp inc2eow inc2exp map2comp map2eow map2exp map2reu man2comp man2eow man2exp nav2comp nav2eow nav2exp trv2comp trv2eow trv2exp trv2pro pm2comp pm2exp pm2pro rm2reu trc2comp trc2eow trc2exp trc2pro comp eow exp ts pro reu}] (prop-mtb) {Maintainability $\eta_5$};
		
		\node[error, above of=prop-mtb] (mtb-error) {$\zeta_5$};
		\path[l] (mtb-error.south west) -- (prop-mtb.45);
		
		\node[property,below of=prop-mtb,actions ocg={}{reset-b bx2pro trv2pro pm2pro trc2pro use2pro ana2pro eco2pro pro}{bx2comp bx2eow bx2exp bx2mtb inc2comp inc2eow inc2exp map2comp map2eow map2exp map2mtb map2reu man2comp man2eow man2exp nav2comp nav2eow nav2exp trv2comp trv2eow trv2exp pm2comp pm2exp rm2reu trc2comp trc2eow trc2exp comp eow exp ts mtb reu}] (prop-pro) {Productivity $\eta_6$};
		
		\node[error, above of=prop-pro] (pro-error) {$\zeta_6$};
		\path[l] (pro-error.south west) -- (prop-pro.45);
		
		\node[property,below of=prop-pro,actions ocg={}{reset-b map2reu rm2reu rep2reu reu}{bx2comp bx2eow bx2exp bx2mtb bx2pro inc2comp inc2eow inc2exp map2comp map2eow map2exp map2mtb man2comp man2eow man2exp nav2comp nav2eow nav2exp trv2comp trv2eow trv2exp trv2pro pm2comp pm2exp pm2pro trc2comp trc2eow trc2exp trc2pro comp eow exp ts mtb pro}] (prop-reu) {Reusability $\eta_7$};
		
		\node[error, above of=prop-reu] (reu-error) {$\zeta_1$};
		\path[l] (reu-error.south west) -- (prop-reu.45);

		
		\begin{scope}[ocg={name=egal,ref=bx2comp,status=visible}]
			\path[l] (factor-bx.east) -- (prop-comp.west);
		\end{scope}
		\begin{scope}[ocg={name=egal,ref=bx2eow,status=visible}]
			\path[l] (factor-bx.east) -- (prop-eow.west);
		\end{scope}
		\begin{scope}[ocg={name=egal,ref=bx2exp,status=visible}]
			\path[l] (factor-bx.east) -- (prop-ex.west);
		\end{scope}
		\begin{scope}[ocg={name=egal,ref=bx2mtb,status=visible}]
			\path[l] (factor-bx.east) -- (prop-mtb.west);
		\end{scope}
		\begin{scope}[ocg={name=egal,ref=bx2pro,status=visible}]
			\path[l] (factor-bx.east) -- (prop-pro.west);
		\end{scope}
		
		\begin{scope}[ocg={name=egal,ref=inc2comp,status=visible}]
			\path[l] (factor-inc.east) -- (prop-comp.west);
		\end{scope}
		\begin{scope}[ocg={name=egal,ref=inc2eow,status=visible}]
			\path[l] (factor-inc.east) -- (prop-eow.west);
		\end{scope}
		\begin{scope}[ocg={name=egal,ref=inc2exp,status=visible}]
			\path[l] (factor-inc.east) -- (prop-ex.west);
		\end{scope}
		
		\begin{scope}[ocg={name=egal,ref=map2comp,status=visible}]
			\path[l] (factor-map.east) -- (prop-comp.west);
		\end{scope}
		\begin{scope}[ocg={name=egal,ref=map2eow,status=visible}]
			\path[l] (factor-map.east) -- (prop-eow.west);
		\end{scope}
		\begin{scope}[ocg={name=egal,ref=map2exp,status=visible}]
			\path[l] (factor-map.east) -- (prop-ex.west);
		\end{scope}
		\begin{scope}[ocg={name=egal,ref=map2mtb,status=visible}]
			\path[l] (factor-map.east) -- (prop-mtb.west);
		\end{scope}
		\begin{scope}[ocg={name=egal,ref=map2reu,status=visible}]
			\path[l] (factor-map.east) -- (prop-reu.west);
		\end{scope}
		
		\begin{scope}[ocg={name=egal,ref=man2comp,status=visible}]
			\path[l] (factor-man.east) -- (prop-comp.west);
		\end{scope}
		\begin{scope}[ocg={name=egal,ref=man2eow,status=visible}]
			\path[l] (factor-man.east) -- (prop-eow.west);
		\end{scope}
		\begin{scope}[ocg={name=egal,ref=man2exp,status=visible}]
			\path[l] (factor-man.east) -- (prop-ex.west);
		\end{scope}
		
		\begin{scope}[ocg={name=egal,ref=nav2comp,status=visible}]
			\path[l] (factor-nav.east) -- (prop-comp.west);
		\end{scope}
		\begin{scope}[ocg={name=egal,ref=nav2eow,status=visible}]
			\path[l] (factor-nav.east) -- (prop-eow.west);
		\end{scope}
		\begin{scope}[ocg={name=egal,ref=nav2exp,status=visible}]
			\path[l] (factor-nav.east) -- (prop-ex.west);
		\end{scope}
		
		\begin{scope}[ocg={name=egal,ref=trv2comp,status=visible}]
			\path[l] (factor-trv.east) -- (prop-comp.west);
		\end{scope}
		\begin{scope}[ocg={name=egal,ref=trv2eow,status=visible}]
			\path[l] (factor-trv.east) -- (prop-eow.west);
		\end{scope}
		\begin{scope}[ocg={name=egal,ref=trv2exp,status=visible}]
			\path[l] (factor-trv.east) -- (prop-ex.west);
		\end{scope}
		\begin{scope}[ocg={name=egal,ref=trv2pro,status=visible}]
			\path[l] (factor-trv.east) -- (prop-pro.west);
		\end{scope}
		
		\begin{scope}[ocg={name=egal,ref=pm2comp,status=visible}]
			\path[l] (factor-pm.east) -- (prop-comp.west);
		\end{scope}
		\begin{scope}[ocg={name=egal,ref=pm2exp,status=visible}]
			\path[l] (factor-pm.east) -- (prop-ex.west);
		\end{scope}
		\begin{scope}[ocg={name=egal,ref=pm2pro,status=visible}]
			\path[l] (factor-pm.east) -- (prop-pro.west);
		\end{scope}
		
		\begin{scope}[ocg={name=egal,ref=rm2reu,status=visible}]
			\path[l] (factor-rm.east) -- (prop-reu.west);
		\end{scope}
		
		\begin{scope}[ocg={name=egal,ref=trc2comp,status=visible}]
			\path[l] (factor-trc.east) -- (prop-comp.west);
		\end{scope}
		\begin{scope}[ocg={name=egal,ref=trc2eow,status=visible}]
			\path[l] (factor-trc.east) -- (prop-eow.west);
		\end{scope}
		\begin{scope}[ocg={name=egal,ref=trc2exp,status=visible}]
			\path[l] (factor-trc.east) -- (prop-ex.west);
		\end{scope}
		\begin{scope}[ocg={name=egal,ref=trc2pro,status=visible}]
			\path[l] (factor-trc.east) -- (prop-pro.west);
		\end{scope}
	
		\begin{scope}[ocg={name=egal,ref=comp,status=visible}]
			\path[l] (mfactor-mms.west) -- (prop-comp.east);
		\end{scope}
		\begin{scope}[ocg={name=egal,ref=eow,status=visible}]
			\path[l] (mfactor-mms.west) -- (prop-eow.east);
		\end{scope}
		
		\begin{scope}[ocg={name=egal,ref=comp,status=visible}]
			\path[l] (mfactor-ms.west) -- (prop-comp.east);
		\end{scope}
		\begin{scope}[ocg={name=egal,ref=eow,status=visible}]
			\path[l] (mfactor-ms.west) -- (prop-eow.east);
		\end{scope}
		
		\begin{scope}[ocg={name=egal,ref=comp,status=visible}]
			\path[l] (mfactor-ts.west) -- (prop-comp.east);
		\end{scope}
		\begin{scope}[ocg={name=egal,ref=eow,status=visible}]
			\path[l] (mfactor-ts.west) -- (prop-eow.east);
		\end{scope}
		
		\begin{scope}[ocg={name=egal,ref=comp,status=visible}]
			\path[l] (mfactor-bus.west) -- (prop-comp.east);
		\end{scope}
		\begin{scope}[ocg={name=egal,ref=eow,status=visible}]
			\path[l] (mfactor-bus.west) -- (prop-eow.east);
		\end{scope}
		\begin{scope}[ocg={name=egal,ref=exp,status=visible}]
			\path[l] (mfactor-bus.west) -- (prop-ex.east);
		\end{scope}
		\begin{scope}[ocg={name=egal,ref=mtb,status=visible}]
			\path[l] (mfactor-bus.west) -- (prop-mtb.east);
		\end{scope}
		\begin{scope}[ocg={name=egal,ref=pro,status=visible}]
			\path[l] (mfactor-bus.west) -- (prop-pro.east);
		\end{scope}
		
		\begin{scope}[ocg={name=egal,ref=comp,status=visible}]
			\path[l] (mfactor-mtl.west) -- (prop-comp.east);
		\end{scope}
		\begin{scope}[ocg={name=egal,ref=eow,status=visible}]
			\path[l] (mfactor-mtl.west) -- (prop-eow.east);
		\end{scope}
		\begin{scope}[ocg={name=egal,ref=exp,status=visible}]
			\path[l] (mfactor-mtl.west) -- (prop-ex.east);
		\end{scope}
		\begin{scope}[ocg={name=egal,ref=ts,status=visible}]
			\path[l] (mfactor-mtl.west) -- (prop-ts.east);
		\end{scope}
		\begin{scope}[ocg={name=egal,ref=mtb,status=visible}]
			\path[l] (mfactor-mtl.west) -- (prop-mtb.east);
		\end{scope}
		\begin{scope}[ocg={name=egal,ref=pro,status=visible}]
			\path[l] (mfactor-mtl.west) -- (prop-pro.east);
		\end{scope}
		\begin{scope}[ocg={name=egal,ref=reu,status=visible}]
			\path[l] (mfactor-mtl.west) -- (prop-reu.east);
		\end{scope}
		
		\begin{scope}[ocg={name=egal,ref=comp,status=visible}]
			\path[l] (mfactor-lsk.west) -- (prop-comp.east);
		\end{scope}
		\begin{scope}[ocg={name=egal,ref=eow,status=visible}]
			\path[l] (mfactor-lsk.west) -- (prop-eow.east);
		\end{scope}
		\begin{scope}[ocg={name=egal,ref=mtb,status=visible}]
			\path[l] (mfactor-lsk.west) -- (prop-mtb.east);
		\end{scope}
		\begin{scope}[ocg={name=egal,ref=pro,status=visible}]
			\path[l] (mfactor-lsk.west) -- (prop-pro.east);
		\end{scope}
		\begin{scope}[ocg={name=egal,ref=reu,status=visible}]
			\path[l] (mfactor-lsk.west) -- (prop-reu.east);
		\end{scope}
		
		\begin{scope}[ocg={name=egal,ref=eow,status=visible}]
			\path[l] (mfactor-exp.west) -- (prop-eow.east);
		\end{scope}
		\begin{scope}[ocg={name=egal,ref=mtb,status=visible}]
			\path[l] (mfactor-exp.west) -- (prop-mtb.east);
		\end{scope}
		
		\begin{scope}[ocg={name=egal,ref=comp,status=visible}]
			\path[l] (mfactor-io.west) -- (prop-comp.east);
		\end{scope}
		\begin{scope}[ocg={name=egal,ref=eow,status=visible}]
			\path[l] (mfactor-io.west) -- (prop-eow.east);
		\end{scope}
		\begin{scope}[ocg={name=egal,ref=pro,status=visible}]
			\path[l] (mfactor-io.west) -- (prop-pro.east);
		\end{scope}
		
		\begin{scope}[ocg={name=egal,ref=eow,status=visible}]
			\path[l] (mfactor-msa.west) -- (prop-eow.east);
		\end{scope}
		
		\begin{scope}[ocg={name=egal,ref=comp,status=visible}]
			\path[l] (mfactor-ius.west) -- (prop-comp.east);
		\end{scope}
		\begin{scope}[ocg={name=egal,ref=eow,status=visible}]
			\path[l] (mfactor-ius.west) -- (prop-eow.east);
		\end{scope}
		\begin{scope}[ocg={name=egal,ref=exp,status=visible}]
			\path[l] (mfactor-ius.west) -- (prop-ex.east);
		\end{scope}
		
		\begin{scope}[ocg={name=egal,ref=reset-b,status=invisible}]
			\node[rectangle,draw,text=red,xshift=3em,yshift=-3em,right of=factor-trc,actions ocg={}{bx2comp bx2eow bx2exp bx2mtb bx2pro inc2comp inc2eow inc2exp map2comp map2eow map2exp map2mtb map2reu man2comp man2eow man2exp nav2comp nav2eow nav2exptrv2comp trv2eow trv2exp trv2pro pm2comp pm2exp pm2pro rm2reu trc2comp trc2eow trc2exp trc2pro comp eow exp ts mtb pro reu}{reset-b}] (reset-b) {Reset};
		\end{scope}
	\end{tikzpicture}
	\caption{Structure model depicting the influence and moderation effects of factors on MTL properties. {\footnotesize(Due to its size, the model has been made interactive using standard PDF features. Clicking one of the exogenous variables $\xi_{1..9}$ will show only influences of the variable as well as all the moderation effects on influenced endogenous variables. Clicking a endogenous variable $\eta_{1..7}$ will show only influences and moderations on the variable. This view can be reset by using the reset button that appears when using the interactive features. To use the interactive features please open the PDF in \textit{Adobe Reader} or \textit{okular}. Other PDF viewers might work but have not been tested.)}}
	\label{fig:structure_model}
\end{figure*}
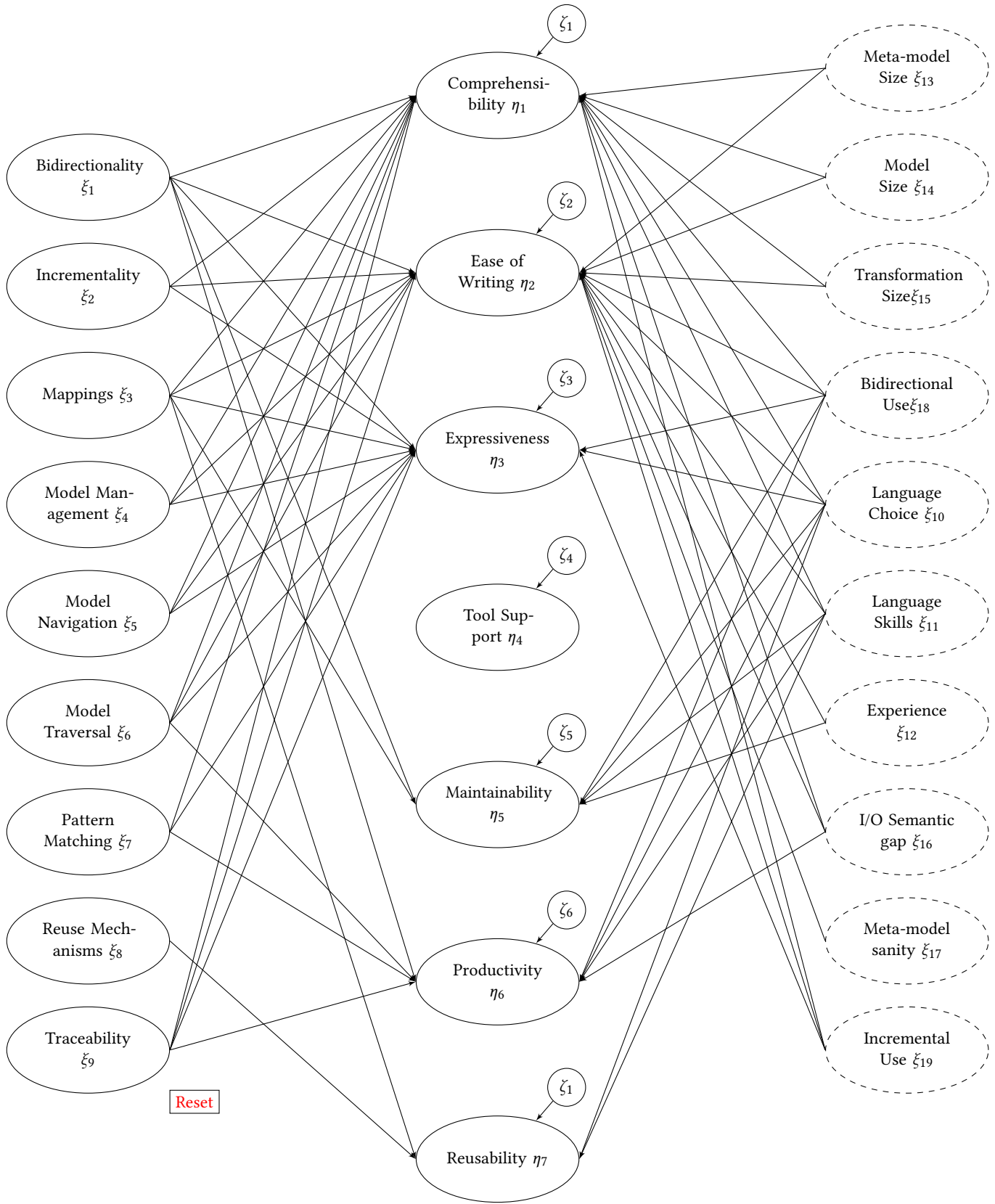

Our study is then guided by the following research questions:

\begin{itemize}
	\item[\textbf{RQ1}] Which of the hypothesised interdependencies withstands a test of significance?
	\item[\textbf{RQ2}] How strong are the influences of model transformation language capabilities on the properties thereof?
	\item[\textbf{RQ3}] How strong are moderation effects expressed by the contextual factors \textit{use-case}, \textit{skills \& experience} and \textit{MTL choice}?
	\item[\textbf{RQ4}] What additional interdependencies arise from the analysis that were not initially hypothesised?
\end{itemize}

As the first study on this subject it contains confirmatory and exploratory elements.
We intend to confirm which of the interdependencies between \textit{MTL capabilities}, \textit{MTL properties} and \textit{contextual properties} extracted from the expert interviews withstand quantitative scrutiny (\textbf{RQ1}).
To do so we want to explore how strong the influence and moderation effects between variables are (\textbf{RQ2 \& RQ3}) to gain new insights and to confirm their significance and relevance (minor influence strengths might suggest irrelevance even if goodness of fit tests confirm a correlation that is not purely accidental).
Lastly, we want to make use of the exploratory elements of USM to find potential interdependencies not hypothesised by the experts in our interviews (\textbf{RQ4}).

\moved{\section{Execution Plan}}
\label{sec:execution_plan}
\rmoved{\ref{review-structure}}{}

Our study is composed of the following steps that will be executed sequentially.

\begin{enumerate}
	\item Development of online survey using an on premise version of the survey tool LimeSurvey\footnote{\url{https://www.limesurvey.org/}}.
	\item Survey review and pilot test by co-authors.
	\item Reworking survey based on pilot test.
	\item Opening online survey to public.
	\item Reaching out to potential survey subjects per mail and social media.
	\item Second round of reaching out (2 weeks later).
	\item Closing of online survey (6 weeks after opening).
	\item Data extraction.
	\item Data analysis using the USM tool NEUSREL\footnote{\url{https://www.neusrel.com}}.
\end{enumerate}
\section{Variables}
\label{sec:vars}

As seen in Figure~\ref{fig:structure_model} there are 26 latent variables relevant to our study.
Variables $\xi_{1..19}$ describe exogenous variables and $\eta_{1..7}$ describe endogenous variables.
Each latent variable is measured through one or more manifest variables.
Here we explain all latent variables and their corresponding manifest variables.
Extending the structure model from Figure~\ref{fig:structure_model} with the manifest variables produces the complete structural equation model evaluated in this study.
Note that USM reduces the requirements for the scale levels of data thus allowing the use of categorical variables in addition to metric variables~\cite{Weiber2021}.

All latent variables related to \textit{MTL capabilities} ($\xi_{1..9}$) are associated with a single manifest variable $x_{1..9}$ which measures in how many of the participants transformations they utilise the capability, using a ratio from 0\% to 100\%.
Similarly, latent variables related to \textit{MTL properties} ($\eta_{1..7}$) are associated with a single manifest variable $y_{1..7}$ which measures the perceived quality of the property on a 5-point likert scale (e.g., very good, good, neither good nor bad, bad, very bad).

The use of single-item scales is a debated topic.
We justify their usage for the described latent variables on multiple grounds.
First, the latent variables are of high complexity due to the abstract concepts they represent.
Second, our study aims to produce first results that need to be investigated in more detail in follow up studies, more focused on single aspects of the model.
And third, due to the size of our structural equation model multi-item scales for all latent variables would increase the size of the survey, potentially putting off many subjects.
The validity of these deliberations for using single-item scales is supported by Fuchs and Diamantopoulos~\cite{fuchs2009using}.

The latent variable \textit{language choice} ($\xi_{10}$) will be measured by means of querying participants to list their 5 most recently used transformation languages and to give an estimate on the percentage of their respective use \% ($x_{10})$.

\textit{Language skills} ($\xi_{11}$) will be measured through $x_{11}$ and $x_{12}$ for which participants are asked to give the amount of years they have been using each language ($x_{11}$) and the amount of hours they use the language per month ($x_{12}$).

Similarly, \textit{experience} ($\xi_{12}$) is associated with the amount of years subjects have been involved in defining model transformations ($x_{13}$) and the amount of hours they spend on developing transformations each month ($x_{14}$).

\textit{Meta-model size} ($\xi_{13}$) and \textit{model size} ($\xi_{14}$) both require participants to state the range between which their (meta-) models vary ($x_{15}$, $x_{16}$).
This is measured by offering participants a number of ranges of (meta-) model objects.
For each range participants should give an estimate on how much percent of the (meta-) models they work fall within that size range.
For models the ranges are: \#objects $\leq 10$, $10 \le$ \#objects $\leq 100$, $100 \le$ \#objects $\leq 1000$, $1000 \le$ \#objects $\leq 10000$, $10000 \le$ \#objects $\leq 100000$, $100000 \le$ \#objects.
For meta-model the ranges are: \#objects $\leq 10$,  $10 \le$ \#objects $\leq 20$, $20 \le$ \#objects $\leq 50$, $50 \le$ \#objects $\leq 100$, $100 \le$ \#objects $\leq 1000$, $1000 \le$ \#objects.
Similarly, \textit{Transformation size} ($\xi_{15}$) is measured on a range of lines of code ($x_{17}$).
The options being: LOC $\leq 100$, $100 \le$ LOC $\leq 500$, $500 \le$ LOC $\leq 1000$, $1000 \le$ LOC $\leq 5000$, $5000 \le$ LOC $\leq 10000$, $10000 \le$ LOC.
Querying size data in this manner and the associated ranges have been successfully applied in a prior work the authors were involved in~\cite{JOT:issue_2021_02/article5}.

To formulate the \textit{semantic gap between input and output} ($\xi_{16}$) we elicit the similarity of the structure ($x_{18}$) and data types ($x_{19}$) on a 5-point likert scale (very similar, similar, neither similar nor dissimilar, dissimilar, very dissimilar).
Participants are asked to give the percentage of all their meta-models that fall within each of the five assessments.

The \textit{meta-model sanity} ($\xi_{17}$) is measured through means of how well participants perceive their structure ($x_{20}$) and their documentation ($x_{21}$) to be on a 5-point scale (very well, well, neither well nor bad, bad, very bad).
Participants are asked to give the percentage of all their meta-models that fall within each of the five assessments.

Lastly, for both \textit{bidirectional uses} ($\xi_{18}$) and \textit{incremental uses} ($\xi_{19}$) we query participants on the ratio of bidirectional ($x_{22}$) and incremental ($x_{23}$) transformations compared to simple uni-directional transformations they have written.
\added{\section{Survey Structure}\label{sec:survey_plan}}

\radded{\ref{review-survey-details} \& \ref{review-survey-details2}}{}

\added{We plan our online survey as a \textit{structured} survey.
There is no need for open questions, because our aim is to quantitatively asses the interdependencies between variables.

}
\moved{The online survey itself will be designed around the following question groups which group related questions together.
Their order follows the suggestions by~\citeauthor{kasunic2005designing}~\cite{kasunic2005designing} to start with easy to answer questions and move towards questions that require subjects to think more about their responses.

\begin{enumerate}
	\item Information and consent about data usage (required for further progression).
	\item Questions on MTL capability usage (\textit{$\xi_{1..9}$}, measured through \textit{$x_{1..9}$}).
	\item Questions about perceived quality properties of transformations and MTLs (\textit{$\eta_{1..7}$}, measured through \textit{$y_{1..7}$}).
	\item Demographic questions (\textit{$\xi_{10..16}$}, measured through \textit{$x_{10..23}$}).
\end{enumerate}

The data in the survey is collected on the basis of the manifest variables ($y_{1..7}$ \& $x_{1..23}$) defined in Section~\ref{sec:vars}.}
\added{
The survey contains one question for each manifest variable.
In total there will be 30 questions spread across the three sections and we expect subjects to require between 35 and 40 minutes on average to complete the survey.
Table~\ref{tbl:questions} shows the tentative questions associated with each manifest variable.
The exact measures taken for each manifest variable are described in Section~\ref{sec:vars}.

Questions within each group follow the same style.
This way participants can familiarise themselves with the question style and can get more efficient answering them.
In this way, subjects can quickly familiarise themselves with the questions in each group.

The questions are designed based on guidelines from \citeauthor{kasunic2005designing}~\cite{kasunic2005designing} and \citeauthor{7968166}~\cite{7968166}.
Questions are \textit{short}, \textit{concrete}, use \textit{simple words} and no \textit{hypotheticals}.

The MTL capability terms can have multiple, slightly different, interpretations within the community.
To prevent ambiguity we provide a short description for each of them before the respective questions.
}

\begin{table*}
        \caption{\added{Survey questions and associated manifest variables in order of appearance in the survey}}~\label{tbl:questions}
	\begin{tabularx}{\linewidth}{@{}Xl@{}}
		\toprule
		\textbf{Question} & \textbf{Manifest Variable}\\
		\midrule
		Which percentage of your transformations utilise \textit{Bidirectionality} functionality? & $x_1$\\
		Which percentage of your transformations utilise \textit{Incrementality} functionality? & $x_2$\\
		Which percentage of your transformations utilise \textit{Mapping} functionality? & $x_3$\\
		Which percentage of your transformations utilise \textit{Model Management} functionality? & $x_4$\\
		Which percentage of your transformations utilise \textit{Model Navigation} functionality? & $x_5$\\
		Which percentage of your transformations utilise \textit{Model Traversal} functionality? & $x_6$\\
		Which percentage of your transformations utilise \textit{Pattern Matching} functionality? & $x_7$\\
		Which percentage of your transformations utilise \textit{Reuse} functionality? & $x_8$\\
		Which percentage of your transformations utilise \textit{Tracing} functionality? & $x_9$\\
		How \textit{comprehensible} are your transformations? & $y_1$\\
		How easy is it to \textit{write} transformations? & $y_2$\\
		How \textit{expressive} are the languages you use? & $y_3$\\
		How is the \textit{tool support} for the languages you use? & $y_4$\\
		How \textit{maintainable} are your transformations?  & $y_5$\\
		How \textit{productive} are you when developing transformations? & $y_6$\\
		How \textit{reusable} are transformation(-parts) in the languages you use? & $y_7$\\
		What are your 5 most recently used languages for writing transformations? & $x_{10}$\\
		How many years have you used your languages? & $x_{11}$\\
		How many hours do you use each language per month? & $x_{12}$\\
		How many years have you worked on defining model transformations? & $x_{13}$\\
		How many hours do you work on writing transformations per month? & $x_{14}$\\
		What number of meta-model elements are used in your transformations? & $x_{15}$\\
		What number of model elements are used in your transformations? & $x_{16}$\\
		How long, measured in LOC, are your transformations? & $x_{17}$\\
		How similar are input and output meta-models in your transformations? & $x_{18}$\\
		How similar are the data types in input and output in your transformations? & $x_{19}$\\
		How well structured are the meta-models in your transformations? & $x_{20}$\\
		How well documented are the meta-models in your transformations? & $x_{21}$\\
		What percentage of your transformations are bidirectional? &$x_{22}$\\
		What percentage of your transformations are incremental? & $x_{23}$\\
		\bottomrule
	\end{tabularx}
\end{table*}

\section{Subjects}
\label{sec:subjects} 

The target subjects are both researchers and professionals from industry that have used dedicated model transformation languages to develop model transformations in the last five years.
We use a combination of convenience sampling and snowball sampling to select our study participants.
Both authors contact researchers and professionals with whom they have personal contact via mail and request them to fill out the online survey.
We further reach out, via mail, to all authors of publications listed in \textit{ACM Digital Library}, \textit{IEEE Xplore}, \textit{Springer Link} and \textit{Web of Science} that contain the key word \textit{model transformation} from the last five years.
A third source of subjects is drawn from social media.
The authors use their available social media channels to recruit further subjects by posting about the online-survey on the platforms.
All participants are requested to forward the online survey to people from their network that they think could also provide useful insights.
\radded{\ref{review-recruiting}}{Participation is voluntary and we do not plan to pay subjects to participate.
This decision is rooted in our experience in previous studies one other survey with 83 subjects \cite{JOT:issue_2021_02/article5} and the interview study we are basing this study on with 56 subjects \cite{Hoeppner2022}.
}

Our intention behind the snowballing is to reach a larger number of industry participants which are hard to reach using a systematic approach based on publications.

It is suggested in literature to have between 5 to 10 times as many participants as the largest number of parameters to be estimated in each structural equation (i.e., the largest number of incoming paths for a latent model variable)~\cite{buckler2008identifying}.
\modified{Thus, the minimal number of subjects for our study to achieve stable results is 80.}
To gain any meaningful results a sample size of 30 must not be undercut~\cite{buckler2008identifying}.
\section{Analysis Plan}
\label{sec:analysis_plan}

\radded{\ref{review-survey-details}}{We intend to execute our analysis regardless of the final sample size but will critically assess the results if it undercuts the thresholds of 80 and 30 respectively.}

We plan on using USM to examine the hypotheses system modelled by the structure model shown in Figure~\ref{fig:structure_model}.
USM requires a declaration of \radded{\ref{review-difference-interviews}}{an initial} likelihood of an interdependence between two variables.
\radded{\ref{review-difference-interviews}}{This is used as a starting points for calculating influence weights but can change over the course of calculation.}
For this, Buckler et al.~\cite{buckler2008identifying} suggest to only assign a value of 0 to those relationships that are known to be wrong.
We use the results of our interview study~\cite{Hoeppner2022}, shown in the structure model, to assign these values.
For each path that is present in the model we will assume a likelihood of 100\%.
To check for interdependencies that might have been missed by interview participants we will use a likelihood of 50\% for all missing paths between $\xi_{1..19}$ and $\eta_{1..7}$.
This is a compromise between exclusion and inclusion, as the interview results suggest that there is no connection\modified{, but not yet certain}.

The tool NEUSREL will then be used on the extracted empirical data and the described additional input to estimate path weights and moderation weights within the extended structure model\radded{\ref{review-rq4-analysis}}{, i.e. the structure model where each exogenous latent variable is connected to all endogenous latent variables}.
\rmoved{\ref{review-significance-tests} \& \ref{review-significance-tests2}}{
It also runs significance tests via a bootstrapping routine~\cite{buckler2008identifying,mooney1993bootstrapping} and produces the significance values for each influence.
}
The following procedures will then be followed to answer the research questions from Section~\ref{sec:rqs}.

\textbf{RQ1}.
We will reject all hypothesised influences\radded{\ref{review-rq4-analysis}}{, i.e. those present in our structure model in Figure~\ref{fig:structure_model},} that do not pass the statistical significance test, meaning the path coefficient between two (or three in case of moderations) latent variables produced by NEUSREL does not exhibit a high level of significance.
Furthermore we will also reject hypothesised influences where the path coefficient is several magnitudes lower than the median influence of all coefficients.
We assume that such low influences suggest irrelevance.

\textbf{RQ2 \& RQ3}.
All path coefficients produced that were not rejected in \textbf{RQ1} will then provide direct values for the influence and moderation strengths to answer \textbf{RQ2} \& \textbf{RQ3}.

\textbf{RQ4}.
\rmodified{\ref{review-rq4-analysis}}{The same significance criteria we applied to all hypothesised influences for \textbf{RQ1}, we also apply to the extended influences, i.e. those added as potential influences with a likelihood of 50\%.
Those influences that pass the significance test are added to the initial structural model as newly discovered influences.}
\added{\section{Threats to Validity}}
\label{sec:threats}
\radded{\ref{review-threats} \& \ref{review-threats2}}{}

\added{Our study is carefully designed and follows standard procedures for this type of study.
There are, however still threats to validity that stem from design decisions and limitations.
In this section we discuss these threats.}

\added{\subsection{Internal Validity}}
\added{Internal validity is threatened by manual errors and biases of the involved researchers throughout the process.

The two activities where such errors and biases can be introduced are the subject selection and question creation.
The selection criteria for study subjects is designed in such a way, that no ambiguities exist during selection.
This prevents researcher bias.

For the survey questions we aim to only use neutral questions and run a pilot test to ensure this.}

\added{\subsection{External Validity}}
\added{External validity is threatened by our subject sampling strategy and the limitations on the survey questions imposed by the complexity of the subject matter.

We utilise convenience and snowball sampling.
Convenience sampling can limit how representative the final group of interviewees is.
Snowball sampling based on this can strengthen this effect.
Since we do not know the target populations makeup it is difficult to asses the extend of this problem.

Using research articles as a starting point introduces a bias towards researchers.
There is little potential to mitigate this problem during the study design, because
there exists no systematic way to find industry users.

Due to the complexity and abstractness of the concepts under investigation, a measurement via reflective of formative indicators is not possible.
Instead we use single item questions.
We further assume that positive and negative effects of a feature are more prominent if the feature is used more frequently.
This can have a negative effect on the external validity of our results.
However, we consciously decided for these limitations to be able to create a study that concerns itself with all factors and influences at once.}

\added{\subsection{Construct Validity}}
\added{Construct validity is threatened by inappropriate methods used for the study.

Using the results of online surveys as input for structural equation modelling techniques is common practice in market research~\cite{Weiber2021}.
It is less common in computer science.
However, we argue that for the purpose of our study it is an appropriate methodology.
This is because the goal of extracting influence strengths and moderation effects of factors on different properties aligns with the goals of market research studies that employ structural equation modelling.}

\added{\subsection{Conclusion Validity}}
\added{Conclusion validity is mainly threatened by biases of our survey participants.
	
It is possible that people who do research on model transformation languages or use them for a long time are more likely to see them in a positive light.
As such there is the risk that too little experiences will be reported on in our survey.
However, this problem did not present itself in a previous study by us on the subject matter \cite{Hoeppner2022}.
In fact researchers were far more critical in dealing with the subject.
AS a result, there might be a slight positive bias in the survey responses, but we believe this to be negligible.
}
\moved{\section{Related Work}}
\label{sec:rw}
\rmoved{\ref{review-structure}}{}

There are numerous works that explore the possibilities gained through the usage of MTLs such as automatic parallelisation~\cite{9146715,biermann2010lifting,benelallam2015distributed}, verification~\cite{lano2015framework,Ko2015} or simply the application of difficult transformations~\cite{10.1007/978-3-540-75209-7_30}.

There is, however, only a small amount of works trying to evaluate the languages to gain insights into where specific advantages or disadvantages associated with the use of MTLs originate from.
Hebig et al.~\cite{Hebig2018} report on a study to evaluate how the use of different languages, namely ATL, QVT-O and Xtend affects the outcome of students solving several transformation tasks.
Götz et al. published a study on how much complexity stems from what parts of ATL transformations~\cite{gotz2020investigating} and compared these results with data for transformations written in Java~\cite{Hoeppner2021} to elicit advantageous features in ATL and to explore what use-cases justify the use of a general purpose language over a model transformation language.
Lastly there is our interview study~\cite{Hoeppner2022} on which this registered report is based on.
The interviews brought forth insights into factors from which the advantages and disadvantages of MTLs originate from as well as suggested a number of moderation effects on the effects of these factors.
Apart from these studies there is little effort put into empirical investigation into model transformation languages as shown by Götz et al.~\cite{Goetz2020}.



\bibliographystyle{ACM-Reference-Format}
\bibliography{MTL_survey}

\appendix
\added{\section{MTL capability and property overview}\label{sec:descriptions}}

For short descriptions of all MTL capabilities see Table~\ref{tbl:cap_descriptions}.
For short descriptions of all MTL properties see Table~\ref{tbl:prop_descriptions}.
More thorough explanations can be found in our previous works \cite{Goetz2020,Hoeppner2022}.

\begin{table}[h]
	\caption{MTL capability descriptions}\label{tbl:cap_descriptions}
	\begin{tabularx} {\linewidth} {@{}lX@{}}
		\toprule
		\textbf{MTL capability} & \textbf{Description}\\
		\midrule
		Bidirectionality & Capabilities to define transformations from source to target and target to source in one transformation rule\\
		Incrementality & Capabilities to define transformations that are applied whenever the source model changes and only manipulate those parts that did change\\
		Mappings & Capabilities to explicitly define correspondence between input and output elements\\
		Model Management & Capabilities to read and write models from and to files\\
		Model Navigation & Capabilities to navigate a given model structure\\
		Model Traversal & Capabilities to automatically find and apply transformations to model elements\\
		Pattern Matching & Capabilities to automatically match patterns of model elements and apply transformations to them\\
		Reuse Mechanisms & Capabilities to reuse whole transformations or parts of transformations\\
		Tracing & Capabilities to (automatically) generate and maintain trace links between source and target elements\\
		
		\bottomrule
	\end{tabularx}
\end{table}

\begin{table}[h]
	\caption{MTL property descriptions}\label{tbl:prop_descriptions}
\begin{tabularx} {\linewidth} {@{}lX@{}}
	\toprule
	\textbf{MTL property} & \textbf{Description}\\
	\midrule
	Comprehensibility & The ease of reading and understanding the purpose and functionality of a transformation\\
	Ease of Writing & The ease at which a developer can produce a transformation for a specific purpose\\
	Expressiveness & The amount of dedicated transformation concepts in a language\\
	Productivity & The degree of effectiveness and efficiency with which transformations can be developed and used\\
	Maintainability & The degree of effectiveness and efficiency with which a transformation can be modified\\
	Reusability & The ease of reusing transformations or parts of transformations to create new transformations (with different purposes)\\
	Tool Support & The amount of tools that exist to support developers in their efforts\\
	
	\bottomrule
\end{tabularx}
\end{table}

\end{document}